\documentclass[sigconf,screen]{acmart}

\AtBeginDocument{%
  }

\acmYear{2026}\copyrightyear{2026}
\setcopyright{cc}
\setcctype[4.0]{by}
\acmConference[FSE Companion '26]{34th ACM Joint European Software Engineering Conference and Symposium on the Foundations of Software Engineering}{July 5--9, 2026}{Montreal, QC, Canada}
\acmBooktitle{34th ACM Joint European Software Engineering Conference and Symposium on the Foundations of Software Engineering (FSE Companion '26), July 5--9, 2026, Montreal, QC, Canada}
\acmDOI{10.1145/3803437.3805207}
\acmISBN{979-8-4007-2636-1/26/07}

\usepackage{tabularray}
\usepackage{orcidlink}
\usepackage{csquotes}
\usepackage{breakcites}
\usepackage{enumitem}
\usepackage{balance}
\usepackage{glossaries}
\glsdisablehyper
\setacronymstyle{long-short}

\newcommand{\glsIfFirst}[3]{%
  \ifglsused{#1}%
    {\gls{#1}#3} %
    {\gls{#1}#2}%
}

\newacronym{llm}{LLM}{large language model}
\newacronym{ai}{AI}{artificial intelligence}
\newacronym{genai}{GenAI}{generative artificial intelligence}

\newcommand{\secref}[1]{\textit{\nameref*{#1}} (Section~\ref{#1})}

\begin{document}

\title{Adoption of Generative Artificial Intelligence in the German Software Engineering Industry: An Empirical Study}

\author{Ludwig Felder \orcidlink{0009-0007-7854-0778}\,}
\affiliation{%
  \institution{Technical University of Munich}
  \city{Heilbronn}
  \country{Germany}}
\authornote{Also with Heilbronn Data Science Center.}
\authornote{Also with Munich Data Science Institute.}
\email{ludwig.felder@tum.de}

\author{Tobias Eisenreich \orcidlink{0009-0004-7168-251X}}
\affiliation{%
  \institution{Technical University of Munich}
  \city{Heilbronn}
  \country{Germany}}
\email{tobias.eisenreich@tum.de}

\author{Mahsa Fischer \orcidlink{0009-0000-6004-6789}}
\affiliation{%
  \institution{Heilbronn University of Applied Science}
  \city{Heilbronn}
  \country{Germany}}
\email{mahsa.fischer@hs-heilbronn.de}

\author{Stefan Wagner \orcidlink{0000-0002-5256-8429}}
\affiliation{%
  \institution{Technical University of Munich}
  \city{Heilbronn}
  \country{Germany}}
\email{stefan.wagner@tum.de}

\author{Chunyang Chen \orcidlink{0000-0003-2011-9618}\,}
\affiliation{%
  \institution{Technical University of Munich}
  \city{Heilbronn}
  \country{Germany}}
\authornotemark[1]
\authornotemark[2]
\authornote{Also with fortiss GmbH.}
\email{chun-yang.chen@tum.de}
\renewcommand{\shortauthors}{Felder et al.}

\begin{abstract}
    \Gls{genai} tools have seen rapid adoption among software developers. While adoption rates in the industry are rising, the underlying factors influencing the effective use of these tools, including the depth of interaction, organizational constraints, and experience-related considerations, have not been thoroughly investigated. This issue is particularly relevant in environments with stringent regulatory requirements, such as Germany, where practitioners must address the GDPR and the EU AI Act while balancing productivity gains with intellectual property considerations. Despite the significant impact of \gls{genai} on software engineering, to the best of our knowledge, no empirical study has systematically examined the adoption dynamics of \gls{genai} tools within the German context. To address this gap, we present a comprehensive mixed-methods study on \gls{genai} adoption among German software engineers.
    Specifically, we conducted 18 exploratory interviews with practitioners, followed by a developer survey with 109 participants. We analyze patterns of tool adoption, prompting strategies, and organizational factors that influence effectiveness. Our results indicate that experience level moderates the perceived benefits of \gls{genai} tools, and productivity gains are not evenly distributed among developers. Further, organizational size affects both tool selection and the intensity of tool use. Limited awareness of the project context is identified as the most significant barrier. We summarize a set of actionable implications for developers, organizations, and tool vendors seeking to advance \glsIfFirst{ai}{ }{-}assisted software development.

\end{abstract}

\ccsdesc[500]{Software and its engineering~Software development techniques}
\ccsdesc[300]{Software and its engineering~Software development process management}
\ccsdesc[300]{Software and its engineering~Development frameworks and environments}
\ccsdesc[500]{Human-centered computing~Empirical studies in HCI}

\keywords{Generative AI, Large Language Models, Software Engineering, Developer Productivity, Empirical Study, AI-Assisted Development, Code Generation, Human-AI Interaction}

\maketitle
\glsresetall %

\section{Introduction}
The integration of \gls{genai} into software engineering has fundamentally altered the development landscape. \Glspl{llm} have evolved from impressive experiments to daily drivers. Developers now use them throughout the life cycle for code generation, explanation, and testing. Industry surveys estimate over 70\% of developers use \gls{ai} tools in some capacity.\footnote{\url{https://survey.stackoverflow.co/2025/ai}} From conversational interfaces like ChatGPT to IDE assistants, these tools promise productivity gains. However, high adoption rates can mask a more complex reality: having a tool does not guarantee effective use or workflow integration.

Previous studies have primarily examined the extent of \gls{genai} adoption, focusing on which tools are used and how frequently\cite{russo2024navigating}. However, there has been limited investigation into the depth of interaction, including the prompting strategies practitioners use, the organizational constraints that influence adoption, and the experience-related factors that determine whether \gls{ai} tools enhance productivity or introduce new challenges. This gap is important because the difference between theoretical capabilities and practical utility is often increased by the friction of adoption and integration. As developers shift from traditional coding to coordinating \gls{ai} agents, it is necessary to identify and understand these obstacles to establish effective software engineering practices.

In this study, we focus on the German software engineering industry. This context provides a unique opportunity to analyze \gls{ai} adoption, driven by three interconnected factors.
First, German organizations operate under the European Union's General Data Protection Regulation (GDPR)\footnote{\url{https://gdpr.eu/tag/gdpr/}} and Artificial Intelligence Act\footnote{\url{https://eur-lex.europa.eu/legal-content/EN/TXT/?uri=CELEX:32024R1689}}, which impose strict requirements on data processing, storage, and transfer. For software teams, this creates immediate tensions when using cloud-based \gls{ai} services that may process proprietary code or sensitive data on external servers~\cite{groplerFutureGenerativeAI2025}. Unlike markets where regulatory frameworks are still emerging, German practitioners must navigate established compliance requirements that directly affect tool selection and usage patterns. Recent qualitative research confirms these dynamics: ~\citet{neumannPolicyPracticeGenAI2026} found that regulatory pressures in German organizations are often translated into restrictive policies without accounting for actual usage patterns, creating systematic gaps between policy and practice.

Second, Germany's industrial structure is characterized by the \enquote{Mittelstand}: small and medium-sized enterprises (SMEs) that form the backbone of the economy~\cite{pahnkeGermanMittelstandAntithesis2019}. These organizations often lack the resources to implement private \gls{ai} infrastructure, yet they still face the same competitive pressures that drive \gls{ai} adoption worldwide. This situation raises an important question. How do organizations with limited resources balance the potential productivity benefits of \gls{genai} with compliance requirements and concerns about intellectual property?

Third, the engineering culture in Germany emphasizes quality, comprehensive documentation, and systematic processes~\cite{GermanStandardizationRoadmap}. This cultural background may lead practitioners to approach \glsIfFirst{ai}{ }{-}generated code with greater skepticism and to apply more rigorous validation practices than in environments that prioritize rapid deployment. Studying adoption and usage in this context helps to identify challenges that may not be addressed in other markets.

Prior empirical evidence supports the distinctiveness of this context. Recent global surveys indicate that \gls{ai} tool adoption among German developers lags behind countries such as India, Brazil, and the United States\footnote{\url{https://github.blog/news-insights/research/survey-ai-wave-grows/}}. However, lower adoption rates alone do not explain the underlying dynamics. Are German developers more cautious due to regulatory constraints? Do organizational policies actively discourage certain tools? Or do experienced practitioners in quality-focused industries perceive less value from current \gls{ai} capabilities? 

To investigate this landscape, we use an exploratory sequential mixed-methods approach. We first conducted 18 semi-structured interviews with practitioners, 
from both large corporations and SMEs, to identify key themes related to \gls{ai} adoption barriers and interaction patterns. Based on these qualitative findings, we developed a comprehensive survey that collected responses from 109 software engineers working mainly in Germany. This approach enables us to move beyond basic adoption statistics and to analyze the practices, perceptions, and contextual factors that influence effective \gls{ai} integration.

Specifically, we address the following research questions:
\begin{itemize}[leftmargin=*]
    \item \textbf{RQ1:} How do software engineers in Germany adopt and interact with \gls{genai} tools in their daily development practices?
    \item \textbf{RQ2:} What challenges impede the effective integration of \gls{genai} into professional software development?
    \item \textbf{RQ3:} How do developer experience and organizational context moderate the perceived effectiveness and adoption of \gls{genai} tools?
\end{itemize}

Our analysis indicates that although \gls{ai} tools are widely adopted, their effectiveness varies considerably. This variation can be attributed to factors such as developer expertise, organizational infrastructure, and the current limitations of \gls{ai} technologies. We identify several important patterns: An Experience Paradox, where junior and senior developers hold differing perspectives; a Context Wall, where the lack of software project-specific knowledge limits its usefulness; and a Corporate Infrastructure Split, where the size of the organization influences tool selection and usage.

The main contributions of our empirical study can be summarized as follows:

\begin{itemize}[leftmargin=*]
    \item We conducted an empirical study of \gls{genai} adoption patterns, prompting strategies, and usage frequencies among German software engineers.
    \item Our analysis of the technical, organizational, and experiential factors that impede or enable effective \gls{ai} integration.
    \item We share implications for individual practitioners, organizations, and tool vendors seeking to improve the effectiveness of \glsIfFirst{ai}{ }{-}assisted software development.
\end{itemize}

The remainder of our paper is organized as follows. Section~2 describes our mixed-method study design. Section~3 presents our results on tool adoption, prompting strategies, and challenges, and Section~4 compares them with the global context. Section~5 discusses key insights and patterns found, and Section~6 synthesizes implications for practice. Section~7 analyzes potential threats and the solutions to mitigate these threats. Section~8 reviews related work on \gls{ai} adoption and usage in software engineering and Section~9 summarizes our empirical study.

\section{Study Design}
\begin{figure*}
    \centering
    \includegraphics[width=1\linewidth]{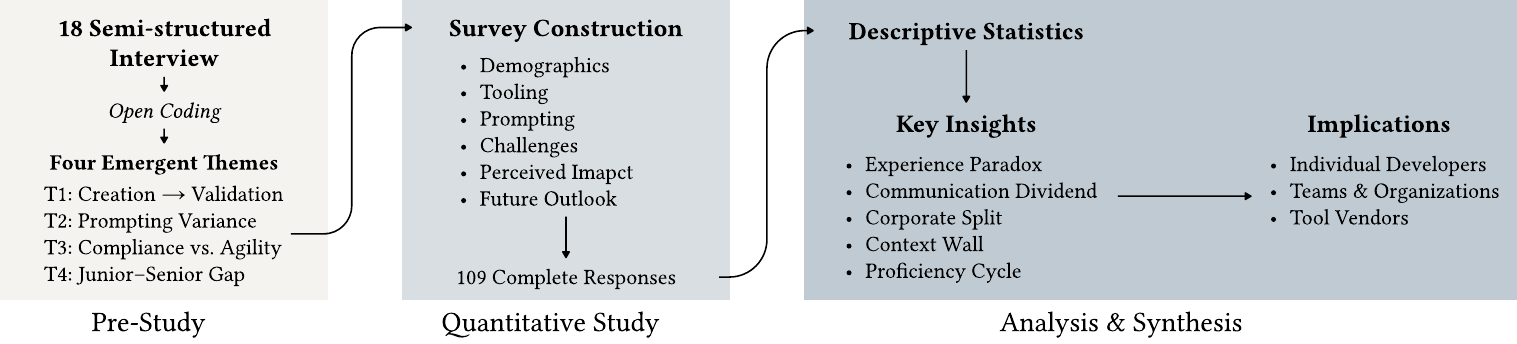}
    \caption{Methodology overview of our sequential empirical study.}
    \Description{A chart visualizing the methodology}
    \label{fig:survey_flow}
\end{figure*}

To reflect current industry realities, we adopted an exploratory sequential mixed-methods design. Our approach started with a qualitative pre-study to identify emerging themes, which subsequently guided the design of our quantitative survey, see Figure~\ref{fig:survey_flow}.

\subsection{Qualitative Pre-Study}
We conducted 18 semi-structured interviews to pinpoint specific barriers to \gls{ai} adoption in the engineering of software. The interview guideline is available in the supplementary data. Our participants included Junior and Senior Developers, Project Managers, and CTOs, representing both major German corporations and various small and medium enterprises (SMEs). This sample ensured we captured perspectives from hands-on code implementation through to the strategic tier of IT management. We recorded and transcribed all sessions. 

To connect our analysis to concrete findings, we used open coding to map specific qualitative themes directly to our quantitative survey construction. Two researchers independently coded all transcripts using open coding. After initial coding, researchers met to compare codes and resolve discrepancies through discussion until consensus was reached.
To connect our analysis to concrete findings, we used open coding to map specific qualitative themes directly to our quantitative survey construction. This qualitative grounding allowed us to identify industrial friction points, challenges, and novel workflows, yielding four primary themes:
\begin{itemize}[leftmargin=*]
    \item \textbf{Theme 1: The Shift from Creation to Validation:} Interviewees consistently reported a transition from writing code to reviewing and debugging \glsIfFirst{ai}{ }{-}generated output. They described a workflow in which the human developer primarily serves as a reviewer of \glsIfFirst{ai}{ }{-}generated logic rather than its sole author.
    \item \textbf{Theme 2: Variance in Prompting:} Our interviews revealed a big variance in prompting strategies. Some relied primarily on basic zero-shot queries, others described a more complex ``Context Engineering'': They reported pasting file trees and database schemas, or using framework-specific tools. This often included multi-turn conversations before instructing the model to generate implementations.
    \item \textbf{Theme 3: Compliance vs. Agility Dilemma:} We observed distinct adoption patterns diverging by organizational size. Participants from larger companies reported strict data sovereignty policies that prohibit the use of public \glspl{llm}, necessitating the use of company internal tools. Conversely, smaller enterprises are concerned about intellectual property leakage but lack the infrastructure to support self-hosting.
    \item \textbf{Theme 4: The Junior-Senior Competency Gap:} Several senior engineers expressed concerns over an erosion of fundamentals: Junior developers relying on \gls{ai} might fail to develop deep problem-solving skills, becoming dependent on tools they cannot validate. Participants predicted a role shift where engineers evolve into agent architects who orchestrate \gls{ai} agents rather than code syntax.
\end{itemize}

\subsection{Data Collection and Sampling}
Building on these qualitative findings, we developed a survey, designed to validate these themes across the wider industry. We systematically transferred our four themes in our survey construction. To address the shift form Creation to Validation, we incorporated questions regarding specific tooling and usage patterns. To quantify the prompting strategies we assessed respondents knowledge and application of distinct prompting strategies. The Compliance vs. Agility dilemma was addressed through detailed items on challenges, tool selection and organizational demographics. Finally, the Competency Gap was explored by collecting robust demographic profiling and items measuring the perceived impact of \gls{ai} on the software development process.
The resulting survey combined closed-ended questions to quantify usage patterns and open-ended questions to capture qualitative insights into the changing role of engineers. Our survey was structured into six distinct blocks to capture a holistic view of the developer experience:

\begin{itemize}[leftmargin=*]
    \item \textbf{Demographics:} We collected control variables including professional role, years of experience, educational background, and company size.
    \item \textbf{Tooling and Usage:} Respondents indicated their adoption of specific \gls{ai} models (e.g., ChatGPT, Claude) and integrated tools (e.g., GitHub Copilot). We measured usage intensity using a 5-point scale (ranging from \enquote{Never} to \enquote{Daily}) across seven core development activities, including code generation, testing, bug fixing, and documentation.
    \item \textbf{Prompting Strategies:} We evaluated the perceived effectiveness of ten distinct prompting techniques (e.g., few-shot prompting, role prompting) on a 5-point scale.
    \item \textbf{Challenges and Customization:} Participants rated the severity of 14 potential challenges (e.g., hallucinations, data privacy) and specific integration and customization needs on a 5-point scale.
    \item \textbf{Impact and Sentiment:} We assessed the perceived impact of \gls{ai} on five process dimensions (e.g., workflow speed, bug fixing efficiency) and included open-ended questions to capture qualitative insights regarding the changing role of software engineers.
    \item \textbf{Future Perspectives}: Participants were asked about anticipated technical advancements in the field (e.g. deeper understanding, improved capabilities). Furthermore, open-ended items invited participants to predict the evolution of the software engineering role and articulate specific concerns.
\end{itemize}

For all questions, participants could answer \enquote{I don't know} to avoid forcing them to answer questions they do not feel competent to answer. While the 5-point scales had textual descriptions, we converted them to numerical data ranging from 1 to 5 for some analysis, and reported means for interpretability. The full survey is available in the supplementary data.

We distributed the derived survey via LimeSurvey and collected answers between 15 April and 20 August 2025. We selected participants through convenience sampling by sharing the questionnaire with industry contacts and on the social media platform LinkedIn. Participation was strictly voluntary, with no incentives offered to respondents. Data collection was performed anonymously to ensure privacy and encourage candid responses regarding organizational challenges. A total of 210 responses were initially recorded. We excluded 101 incomplete responses, most of which had no questions answered at all, but were still recorded by the system. The final dataset consisted of $n=109$ complete submissions.

The final sample represents a highly experienced cohort. The majority of respondents are Software Developers and Engineers ($62\%$), followed by Team Lead and Managers ($13\%$), and Software Architects and Requirements Engineers ($8\%$). Most participants possess significant professional experience, with a mean of 12.1 years and a median of 10 years. The sample is skewed towards seniority, with 26\% categorized as Senior (15+ years experience), while Mid-level (5-15 years) and Junior (0-3 years) professionals account for 54\% and 20\% respectively. Respondents represent a broad spectrum of organizational sizes. Medium-sized enterprises (100-999 employees) constitute 29\% of the sample, followed closely by large enterprises (1,000-9,999 employees) at 27\% and large corporations ($>$10,000 employees) at 24\%. Small enterprises (10-99 employees) make up 14\%. With the rest of the responses working in enterprises with less than 10 employees or preferring not to say. Geographically, the study is strongly focused on the German market, with 88\% of respondents working primarily in this region. This matches our target sampling. The sample is characterized by a high level of formal education. The majority hold a Master's degree or equivalent (51\%), followed by Bachelor's degrees (22\%) and Doctorates (14\%). Only 12\% of respondents did not have an academic education, and the remaining 1\% did not want to disclose their education.

\section{The \glsentryshort{AI} Landscape}
This section reports quantitative results on \gls{genai} adoption among German software engineers. Our analysis addresses \textbf{RQ1}, which concerns adoption and interaction patterns, and \textbf{RQ2}, which focuses on challenges to effective integration. We begin by identifying the tools practitioners have adopted and the frequency of their use across development tasks. Next, we analyze the prompting strategies developers use and assess their perceived effectiveness. We then examine the challenges that limit productive use and identify unmet integration needs. Finally, we evaluate the overall perceived impact on productivity. The survey results are available in the supplementary data. They are redacted to exclude incomplete responses, which were not used in our analysis, and free-text responses, as they could aid in de-anonymizing the respondent.

\subsection{Adoption of Models and Tools}

\begin{figure}
    \centering
    \includegraphics[width=1\linewidth]{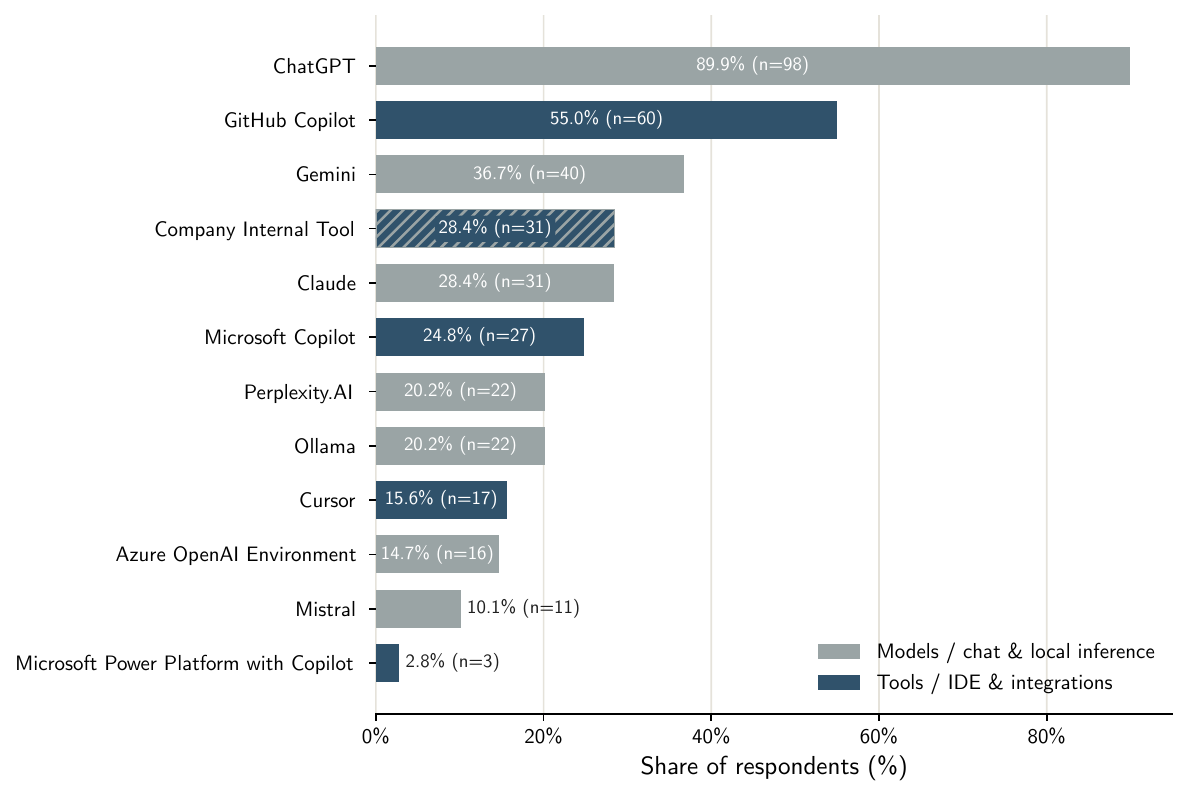}
    \caption{Adoption rates of \gls{genai} models and tools among German software engineers.}
    \Description{A bar chart visualizing the adoption rates of various \gls{genai} tools.}
    \label{fig:tools}
\end{figure}

The adoption of \gls{genai} among German software engineers is split between foundational models with conversational interfaces and integrated tools (see Figure~\ref{fig:tools}). Conversational chats dominate. OpenAI's ChatGPT is used by 90\% of respondents. Google's Gemini follows at 37\%, and Anthropic's Claude at 28\%. Open-source and local inference models are also gaining traction. Ollama, a platform for serving open-source LLMs, has a 20\% usage rate, and Mistral has a 10\% usage rate. Cross-tabulation shows Ollama is used across the industry with the exception of ``large enterprises'' (1.000-9.999) with no adoption at all.
Github's Copilot leads IDE-based solutions with 55\% adoption. A smaller group is using \enquote{\glsIfFirst{ai}{ }{-}native} editors, such as Cursor (16\%). 
Nearly a third (28\%) use internal company tools, though our survey did not distinguish between chat interfaces, IDE integrations, or other deployment forms.

\subsection{Usage Patterns by Task}

\begin{figure}
    \centering
    \includegraphics[width=1\linewidth]{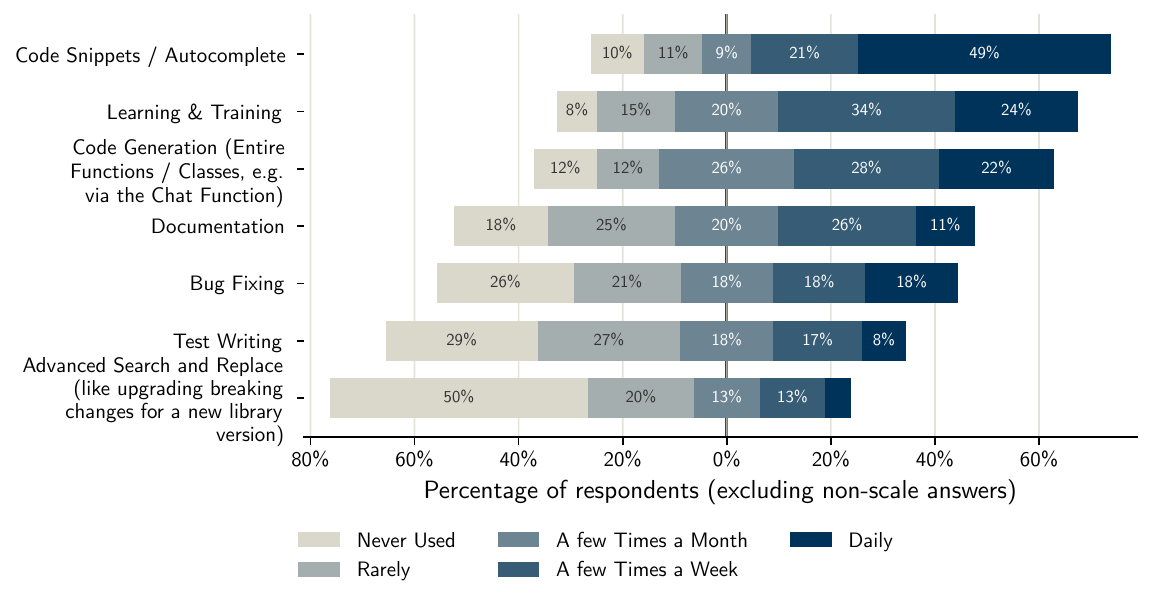}
    \caption{Frequency of \gls{genai} usage across seven software development tasks.}
    \Description{A chart visualizing the \gls{genai} usage profiles of the survey respondents}
    \label{fig:usage}
\end{figure}

Our analysis of usage frequency by task shows a clear divide between creation and validation. \gls{ai} supports generative tasks but is less utilized for quality assurance or structural maintenance (see Figure~\ref{fig:usage}). Code completion and snippet generation show the most common use case, with 70\% of respondents using \gls{ai} at least a few times per week for these tasks. Similarly, it is often used for learning and training, confirming that engineers increasingly leverage \gls{ai} as on-demand tutors. In contrast, validation tasks like bug fixing (used multiple times a week by 36\%) and testing (25\%) have low adoption rates.

The qualitative responses about the anticipated future confirm this role evolution from author to reviewer. Participants described a transition where \textit{``developers will write less code themselves and become coordinators of \gls{ai} programmers''} (P5). They illustrate a fundamental change in daily competencies, as \textit{``instead of writing and testing code, \glsIfFirst{ai}{ }{-}generated code must now be understood, reviewed, and tested''} (P61). This transition is not purely positive, since the offloading of simple generation tasks leaves humans with the cognitively more demanding tasks. The expectation that \gls{ai} frees developers of the hard work might not hold, \textit{``it might be the opposite, and all we will have left are the slow and tedious parts''} (P154).

\subsection{Prompting Strategies}
\label{sec:prompting_strategies}
\begin{figure}
    \centering
    \includegraphics[width=1\linewidth]{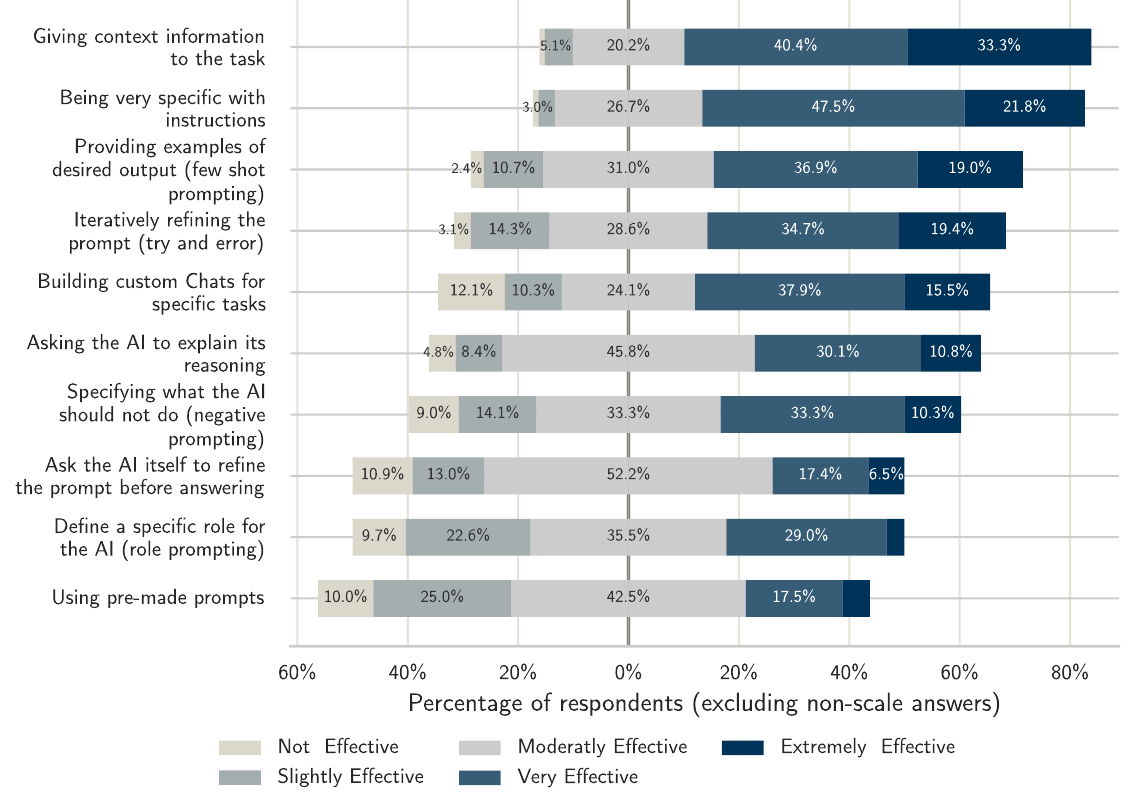}
    \caption{Perceived effectiveness of ten prompting strategies rated on a 5-point scale.}
    \Description{A chart visualizing the perceived effectiveness of various prompting strategies}
    \label{fig:prompting}
\end{figure}

The survey results suggest that developers view \gls{ai} interaction less as a technical engineering task and more as a communication challenge. Strategies that mimic effective human-to-human delegation, specifically \enquote{giving context information} ($Mean=4.0$) and \enquote{being very specific with instructions} ($Mean=3.9$), outperform more sophisticated prompting strategies (see Figure~\ref{fig:prompting}).
Strategies involving iterative refinement fall in the middle tier of effectiveness ratings. \enquote{Providing examples of desired output} (few-shot prompting) achieved a mean of ($Mean=3.6$), followed by \enquote{iteratively refining the prompt through trial and error} ($Mean=3.5$). Notably, the strategy of \enquote{building custom chats for specific tasks} was rated as effective ($Mean=3.3$) but was utilized by a considerably smaller subset of respondents ($n=58$). %
In contrast, advanced techniques that are often popularized in the literature showed poor results in practice: \enquote{Role prompting} (e.g., ``Act as a Senior Architect'') received a mediocre effectiveness score ($Mean=2.9$), while the use of \enquote{pre-made prompt templates} was rated the least effective strategy overall ($Mean=2.8$).

This distinct preference for context over ``prompt hacks'' could be attributed to the limitations of general-purpose models in specialized domains. One engineer noted that standard models struggle with specialized programming languages, expressing a need to \textit{``force-feed an LLM''} (P63) with documentation, but are limited by the current context window size. This suggests that prompting techniques cannot compensate for a model's fundamental lack of domain knowledge. Furthermore, the reliance on \enquote{iterative refinement} is often viewed as a failure mode rather than a feature. The friction of iteration leads some developers to conclude that \textit{``prompting and prompt refining ends up taking longer than writing code just about every time''} (P162), undermining the productivity promises of \gls{ai}. Recent developments, such as the Model Context Protocol (MCP), which became popular during the time our survey closed, are already addressing some of these limitations.

\subsection{Challenges}

\begin{figure}
    \centering
    \includegraphics[width=1\linewidth]{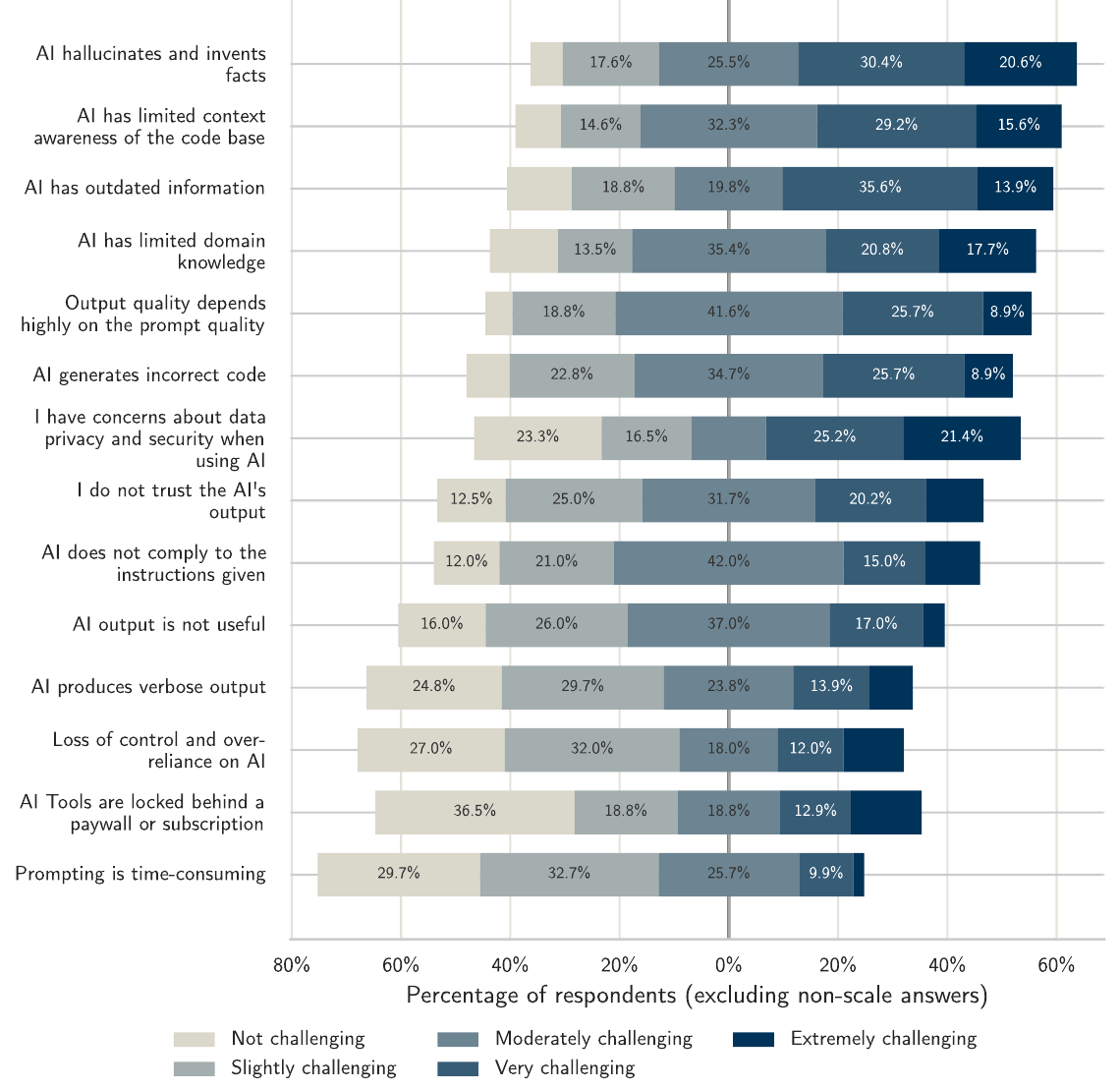}
    \caption{Severity ratings for 14 challenges in applying \gls{genai} to software development.}
    \Description{A chart visualizing the challenges for applying \gls{genai} to software development}
    \label{fig:challenges}
\end{figure}

The primary obstacles to \gls{ai} adoption are related to the fundamental trust gap regarding the reliability and integrity of the model's outputs. ``\gls{ai} hallucinations and invented facts'' emerged as the single most significant challenge, with a mean severity score of 3.4 (on a scale from 1.0 to 5.0). Notably, 51\% of respondents ($n=102$) rated this issue as \enquote{very} or \enquote{extremely} challenging (see Figure~\ref{fig:challenges}). One participant observed that \textit{``in complex projects, hallucinations are all over the place... [the \gls{ai}] focuses too much on fixing symptoms rather than problems''} (P162).
This skepticism is mirrored by deep concerns regarding the long-term degradation of code quality. Respondents did not merely report errors but predicted system risks ahead. One senior engineer explicitly warned of an impending quality crisis, stating, \textit{``I expect a large amount of low-quality code in the near future''} (P153). Another highlighted the dangerous feedback loops introduced by automated workflows, noting that the situation becomes \textit{``very critical, especially when \glsIfFirst{ai}{ }{-}generated software is also tested with \glsIfFirst{ai}{ }{-}generated tests''} (P199).

These challenges are amplified by the \enquote{limited context awareness of the codebase} ($Mean=3.3$) and \enquote{outdated information} ($Mean=3.2$), traits which reduce the quality of generated artifacts. The knowledge cutoff is particularly acute, with 50\% of respondents flagging the model's knowledge cutoff as a critical issue. In complex codebases, the lack of context leads models to fail at root-cause analysis.

\enquote{Data privacy and security concerns} also rank in the top tier of challenges ($Mean=3.1$), with 47\% of respondents rating it as a high-severity issue, underscoring the tension between cloud-based inference and intellectual property protection. One participant expressed concerns that \textit{``\gls{ai} in most applications these days is a service, not a product, it phones home anything it claims to need for its server-side processing''} (P69).

Interestingly, challenges that could be mitigated by more advanced prompting strategies, such as \enquote{The \gls{ai} does not comply to the instructions given}, \enquote{the \gls{ai} output is not useful}, \enquote{the \gls{ai} produces verbose output}, and \enquote{prompting is time consuming} rank among the least relevant challenges in our survey.

\subsection{Integration and Customization Needs}
\label{sec:integration}

\begin{figure}
    \centering
    \includegraphics[width=1\linewidth]{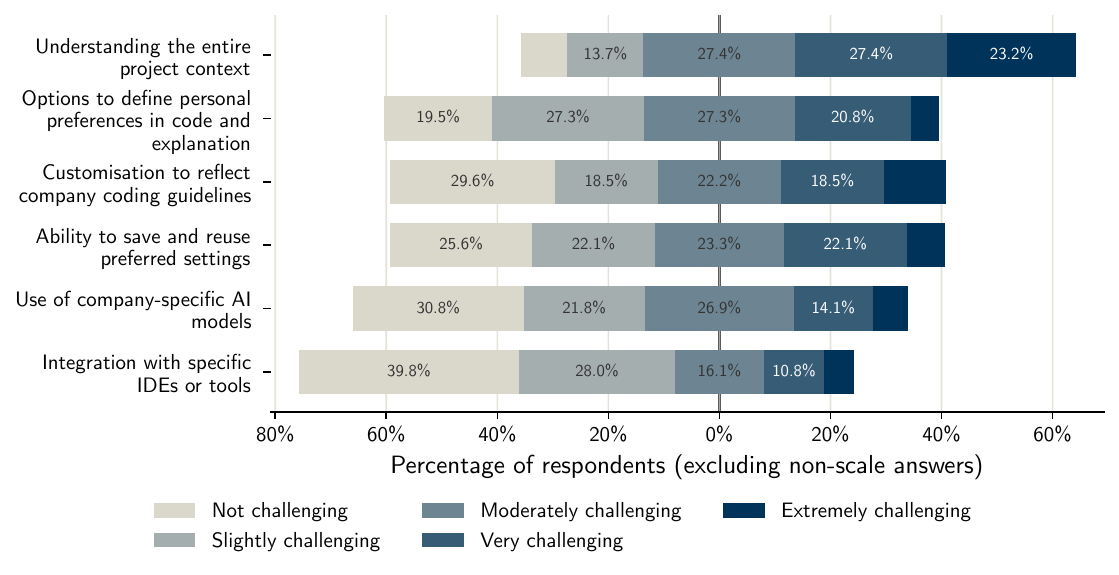}
    \caption{Severity of integration and customization challenges for \gls{genai} tools.}
    \Description{A chart visualizing the challenges of customizing and integrating \gls{genai} tools}
    \label{fig:customization}
\end{figure}

As adoption increases, user friction shifted. Integration and customization are emerging challenges. More than half of the respondents rate \gls{ai}'s inability to grasp full project context as \enquote{Very} or \enquote{Extremely} challenging (see Figure~\ref{fig:customization}), surpassing concerns, such as \enquote{reflecting company coding guidelines} ($Mean=2.6$) or \enquote{integration with specific IDEs} ($Mean=2.1$).
Responses clarify that this is a structural limitation of \gls{ai} agents, not just a user interface issue. One respondent observed that \textit{``current \gls{ai} assistants... just see part of the code and miss dependencies to functions outside of respective files''} (P179). This lack of awareness forces developers to manually bridge the gap between the model's local inference and the broader system architecture, effectively limiting the utility of \gls{ai} to a subset of files, rather than the full project scope. 

\subsection{Perceived Impact}

\begin{figure}
    \centering
    \includegraphics[width=1\linewidth]{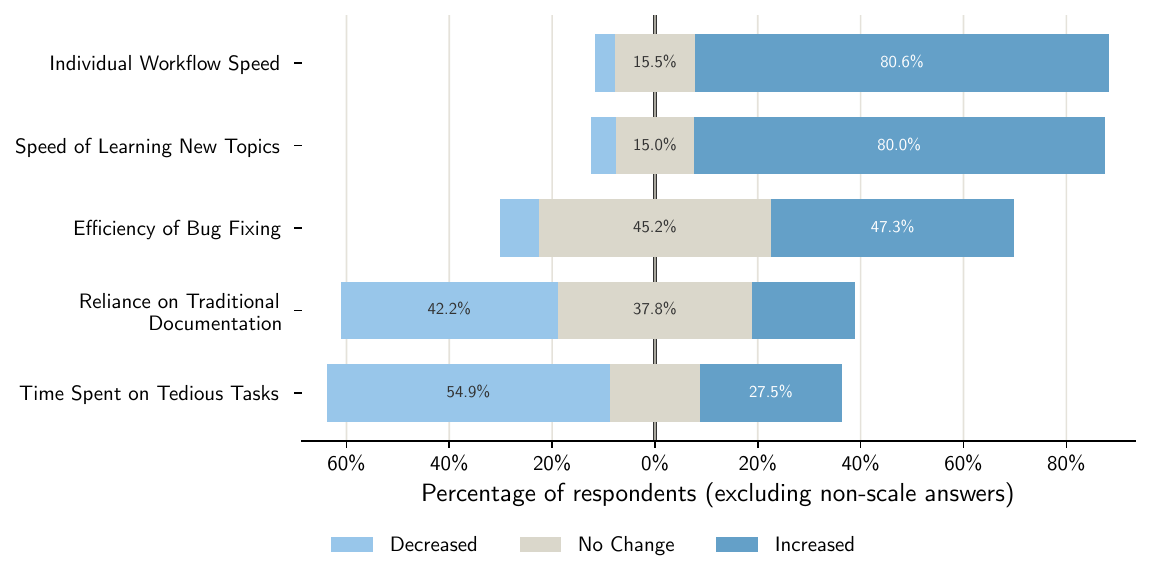}
    \caption{Perceived impact of \gls{genai} on five dimensions of the software development process. }
    \Description{A chart visualizing the perceived change in workflow speed by the adoption of \gls{genai} tools}
    \label{fig:impact}
\end{figure}

Despite the structural limitations discussed above, the perceived impact on personal productivity is overwhelmingly positive, see Figure~\ref{fig:impact}. A striking 76\% of respondents report an increase in their \textit{Individual Workflow Speed}, and 73\% report a faster \textit{Speed of Learning New Topics}. This aligns with the view on \gls{ai} as a \enquote{force multiplier} that allows companies to \textit{``produce more value at a higher pace using software engineers as a foundation''} (P106).
However, the impact on quality assurance and maintenance is more ambiguous. Only 40\% report an improvement in their \textit{efficiency of Bug Fixing}, with 39\% perceiving no change. One respondent offered an explanation, noting that  \textit{``bug fixing takes longer than before, because you are not fixing your own code''} (P61). The cognitive load shifts from creation to reverse-engineering \glsIfFirst{ai}{ }{-}generated logic because the code often lacks the structure a developer would use.

Furthermore, there is a distinct fear that reliance on these tools erodes fundamental competencies. Respondents warned that \textit{``with continuous and extensive use of \gls{ai}, the basic understanding of how the code works is lost.''}, leading to a future where \enquote{\textit{software developers without \gls{ai} tools are no longer able to write programs}} (P201). We remark a growing tension between short-term efficiency gains and the fear about the long-term sustainability of engineering expertise.

\section{Global Context}
We compared our German sample to the global results from the 2025 Stack Overflow Developer Survey\footnote{\url{https://survey.stackoverflow.co/2025/ai/}}. Because of different question configurations, potential selection biases in only one survey, and similar differences, we do not compare the results directly. However, we can still compare trends and triangulate our results.

The Stack Overflow Developer Survey confirms the tendency for developers to use \gls{ai} tools daily. Nearly half of their participants use \gls{ai} tools daily, with roughly a third using them less often, and a fifth not using them at all. In our results (see Figure~\ref{fig:usage}), we observe a similar distribution, though with very few respondents who do not use \gls{ai} at all.

Furthermore, the Stack Overflow Developer Survey confirms the trend of a changing work environment, with over half of the respondents reporting a positive effect on their productivity, and, additionally, 7\% acknowledging a change in their work due to \gls{ai} tools.

Comparing tool adoption rates, we observe very similar results: ChatGPT is adopted at very high levels by both survey populations, followed by GitHub Copilot, Gemini, Claude, Microsoft Copilot, and Perplexity. The category of \enquote{company internal tools} was not measured in the Stack Overflow Survey. Ignoring it, the ordering of our results matches.

Both surveys agree that hallucinations are the biggest challenge with \gls{ai} agents. However, while the Stack Overflow respondents had strong concerns about the security and data privacy when using \gls{ai} agents, our results suggest otherwise. We assume that this discrepancy is either based on the different question framing or due to a different population selection. We could, however, distinguish two groups in our data: participants who were very concerned about their data security and privacy, and those who were not, with a very low number of indifferent respondents (see Figure~\ref{fig:challenges}).

\section{Key Insights and Patterns}
In this section, we aim to highlight key insights and patterns that emerge beyond the raw results, addressing \textbf{RQ3} on how developer experience and organizational context moderate the perceived effectiveness and adoption of GenAI tools. The following patterns frame the status of \gls{genai} in Germany's industrial software engineering. Further statistical analysis illustrates the tension between tool capability, developer experience, and organizational context.

\subsection{Experience Paradox}
\label{sec:experience-paradox}

\begin{figure}
    \centering
    \includegraphics[width=1\linewidth]{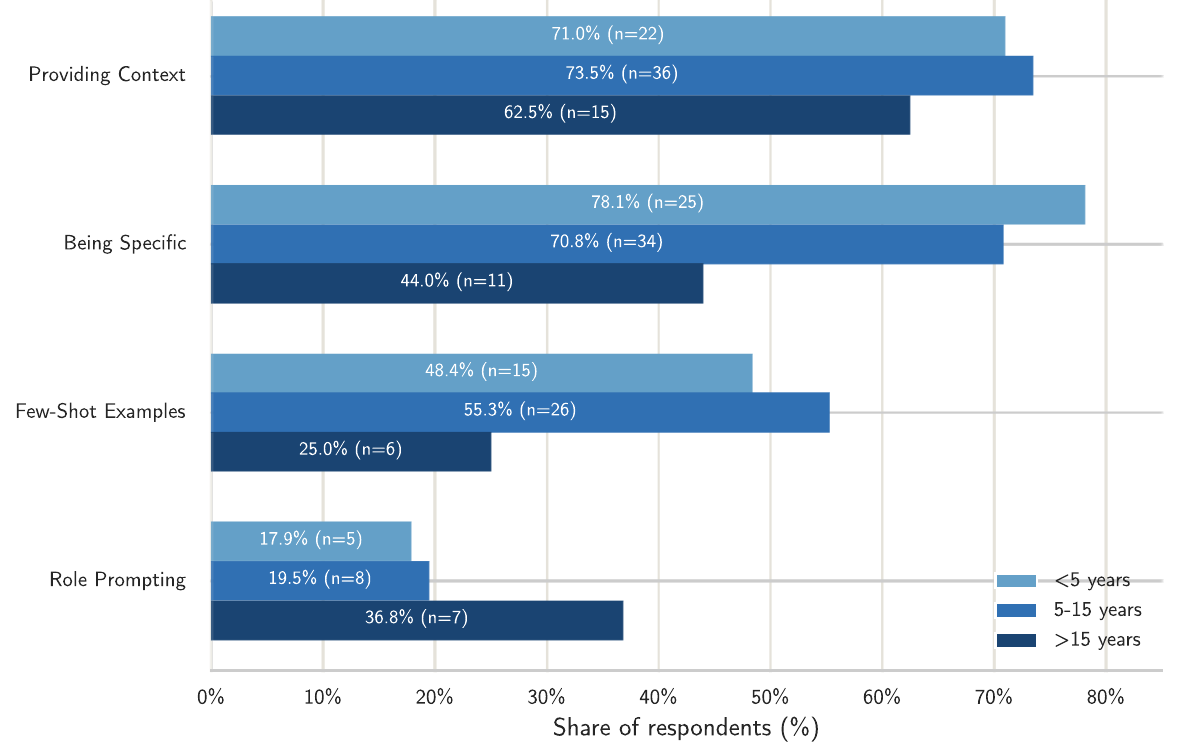}
    \caption{Perceived effectiveness of four prompting strategies stratified by professional experience.}
    \Description{A chart comparing the perceived effectiveness of several prompting strategies to the seniority of survey participants}
    \label{fig:prompting_effectiveness}
\end{figure}

Professional experience seems to have an impact on developers' interaction with \gls{ai}, as revealed by our cross-tabulation analysis. We found a statistically significant relationship between participant experience and the perceived value of specific instructions in prompts ($\chi^2(2, N=109) = 10.84, p_\text{adj} = 0.022$, see Figure~\ref{fig:prompting_effectiveness}): junior engineers ($<5$ years) appreciate specificity (78\% effectiveness), treating \gls{ai} like an oracle, while senior engineers ($>15$ years) are more skeptical (39\% effectiveness). 

Suggesting that greater expertise reduces the perceived utility of specific instruction-following, likely because delegated abstraction exposes the model's reasoning limits, thereby reducing trust among senior staff. This skepticism extends to the assessment of competency itself. Participants expressed concern that \gls{ai} tools mask a lack of understanding, noting that \textit{"we have already seen the rise of vibe coder and I fear that this might become a norm and hard to distinguish between people who are smart or people who use \gls{ai} to appear smart"} (P104). Consequently, rather than leveling the playing field, these tools may worsen existing divides in expertise. As one respondent predicted, \textit{"the bridge between senior engineers with actual core and foundational knowledge of software engineering, and people getting into the software engineering market will grow wider"} (P106). Indicating that the industry may face a difficult transition period as it adapts to these new paradigms.

\subsection{Communication Dividend}
\label{sec:communication_dividend}

\begin{figure}
    \centering
    \includegraphics[width=1\linewidth]{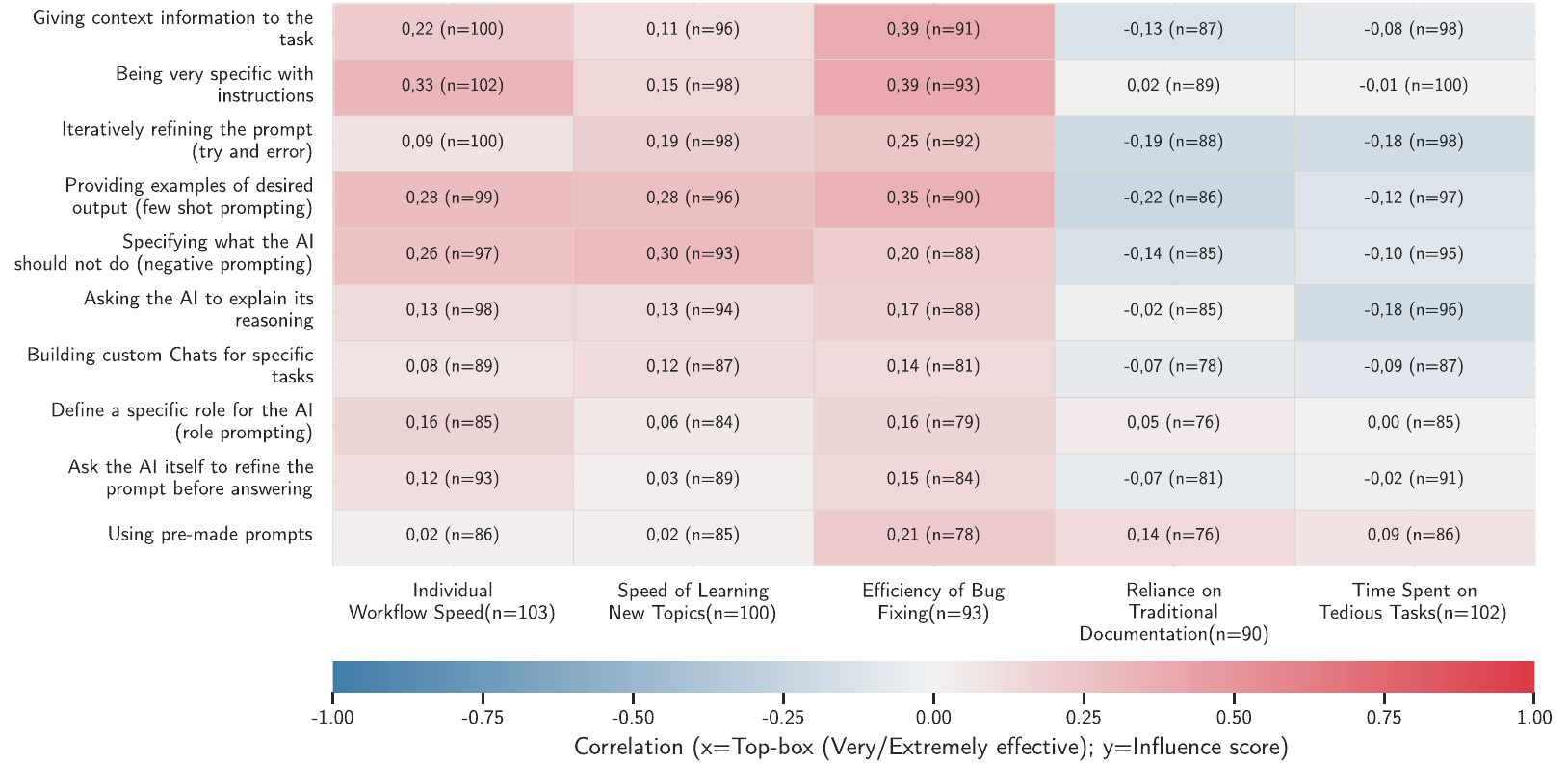}
    \caption{Correlation of effectiveness ratings (very and extremely effective) of prompting strategies and perceived impact.}
    \Description{A chart comparing the perceived effectiveness of prompting strategies with the perceived impact on workflow speed}
    \label{fig:communication}
\end{figure}

Our correlation analysis suggests that effective \gls{ai} interaction may depends more on clear communication than on advanced prompting techniques. As shown in Figure~\ref{fig:communication}, strategies that mirror effective human-to-human delegation yield stronger associations with positive outcomes compared to others. \enquote{Being very specific with instructions} correlates with both workflow speed ($\rho = 0.33$) and bug-fixing efficiency ($\rho = 0.39$), while \enquote{giving context information} shows a similar pattern for bug fixing ($\rho = 0.39$). Few-shot prompting, i.e. providing examples of desired output, also shows correlations across multiple dimensions, including bug fixing ($\rho = 0.35$) and learning ($\rho = 0.28$). In contrast, advanced meta-strategies such as role prompting or using pre-made prompts show weaker associations with perceived impact. These findings suggest that the most productive users do not treat the \gls{ai} as a search engine requiring clever query formulation, but as a collaborator requiring precise scoping and clear intent. However, this relies on the developer's own expertise. As one respondent emphasized, \textit{"you still need to know and understand the domain you're working in. Creating code is one problem, maintaining code is a different problem"} (P116), advocating that the ability to provide clear context is ultimately downstream of fundamental engineering knowledge and experience.

\subsection{Corporate Infrastructure Split}
\label{sec:corp-infra}
Analysis of self-hosted model adoption via Ollama reveals a bimodal distribution: medium enterprises (100–999 employees) and large corporations (10,000+ employees) report similar adoption rates of approximately 35\% (34\% and 35\%, respectively). In contrast, large enterprises (1,000–9,999 employees) report no adoption (0 out of 29 respondents).
This pattern hints at two distinct drivers for self-hosted inference adoption. Large corporations may have the IT resources and data governance requirements that justify investment in private infrastructure. As one respondent noted, while \gls{ai} promises efficiency, organizations are acutely aware that \textit{"we will encounter compliance and IP risks that remain unresolved"} (P150), needing strict control over data flow. In contrast, medium-sized enterprises may adopt Ollama as a lightweight, cost-effective solution to address compliance needs without requiring enterprise-scale infrastructure. 

\begin{figure}
    \centering
    \includegraphics[width=1\linewidth]{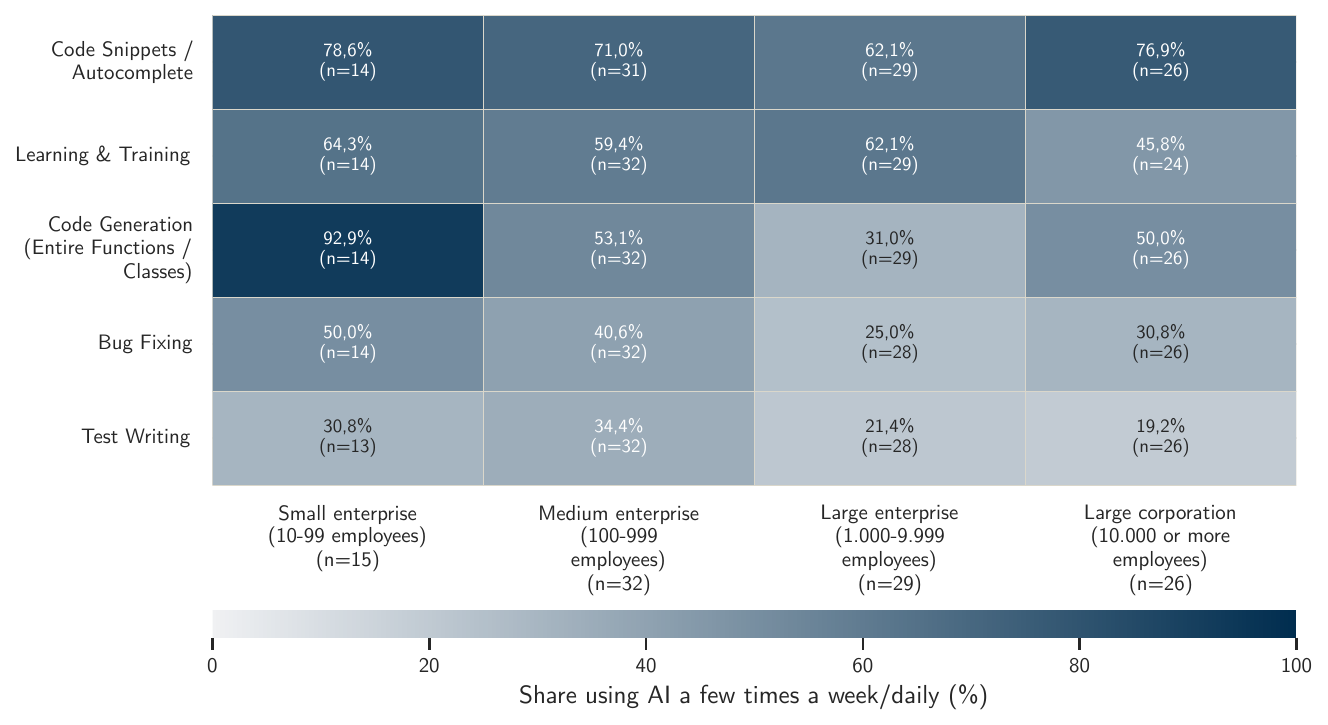}
    \caption{Frequency usage by purpose and company size. Excluding micro-enterprises (<10 employees).}
    \Description{A chart visualizing the usage frequency of \gls{genai} tools by company size}
    \label{fig:usage_company}
\end{figure}

Our results further show that the usage of code generation decreases as company size increases, see Figure~\ref{fig:usage_company}. Small enterprises report the highest rates of weekly or more frequent use (93\%). Indicating a strong focus on leveraging \gls{ai} to maximize productivity. In contrast, only 31-50\% of respondents in large enterprises (1.000-9.999 employees) and large corporation (10.000+) respectively, report similar usage levels.

\subsection{Context Wall}
\label{sec:context-wall}
The most profound limitation identified in our analysis is the \enquote{Context Wall}, in which current models' inability to ground their reasoning in the full reality of a software project, creating a persistent barrier. This limitation can be observed along two dimensions. Spatial context blindness and temporal context decay.
Spatial context blindness refers to the \gls{ai}'s inability to comprehend the holistic structure and interdependencies of a codebase. With 51\% of respondents stating the \gls{ai}'s struggle to understand the entire project context as a severe challenge. 
The severe \enquote{limited context awareness of the codebase} ($Mean=3.3$) and \enquote{limited domain knowledge} ($Mean=3.2$), suggest that \gls{ai} tools struggle to navigate and understand the architecture of complex systems.
Temporal context decay, by contrast, describes the friction caused by the \enquote{knowledge cutoff} inherent to pre-trained models. Nearly half of our respondents (50\%) identify \enquote{outdated information} as a critical issue ($Mean=3.2$). Thereby forcing the users to manually correct deprecated implementations, while generating code that is superficially plausible yet subtly misaligned.

Together, these context gaps impose a measurable productivity cost, a verification tax. Our analysis reveals a negative correlation between perceived workflow speed and distrust in \gls{ai} output ($\rho = -0.33, p < 0.001$). This relationship is consistent with two interpretations: Either the cognitive overhead of validating unreliable outputs may diminish productivity gains, or developers who experience fewer gains may develop greater skepticism toward the tools. This pattern generalizes beyond code generation. Another correlation was found between efficiency of bug fixing and \enquote{distrust in \gls{ai} output} ($\rho = -0.32, p < 0.001$). 
Further, workflow speed correlates with code generation frequency ($\rho = 0.48, p < 0.001$). While heavy adopters report greater acceleration, they also increase their exposure to outputs requiring verification, whether this erodes or reinforces their trust remains an open question.
Our results also hint at emergent mitigation strategies. Notably, among developers using advanced prompting techniques, e.g. asking the \gls{ai} to refine its own prompt before answering, correlates with both workflow speed ($\rho = 0.47, p < 0.01$) and bug fixing efficiency ($\rho = 0.36, p < 0.05$). Importantly, while causality cannot be inferred from these observed correlations, these results indicate that externalizing reasoning steps may partially compensate for the context wall. However, such workarounds place additional meta-cognitive burden on developers and underscore the need for architectural solutions.

\subsection{Proficiency Cycle}\label{sec:proficiency-cycle}
Success in working with \gls{ai} is not evenly distributed. To identify distinct usage profiles, we applied k-means clustering (k=2) on respondents' self-reported usage frequency across our seven \glsIfFirst{ai}{ }{-}assisted development tasks. This revealed two clusters: a moderate-usage group (C0, n=63, 58\%) and a high-usage group (C1, n=46, 42\%), which we term \enquote{power users}. Power users exhibit elevated reliance on \gls{ai} for code generation, autocomplete, bug fixing, and learning support, and report greater perceived productivity benefits.
Demographic analysis reveals that power users tend to have lower levels of education ($\chi^2(5, N=109) = 13.91$, $p_\text{adj} = 0.016$) and work for smaller companies ($\chi^2(4, N=109) = 11.30$, $p_\text{adj} = 0.023$).

Several interpretations may explain this pattern. First, power users may have had comparatively lower baseline performance prior to \gls{genai} adoption, with these tools effectively leveling the field, a hypothesis consistent with the \enquote{Experience Paradox} (Section~\ref{sec:experience-paradox}). Second, power users may work on tasks more amenable to automation, enabling heavier tool reliance. Third, extensive use may itself cultivate more effective interaction strategies, creating a self-reinforcing proficiency cycle. Finally, we cannot rule out response bias: Participants enthusiastic about \gls{genai} may both use tools more frequently and overestimate their productivity impact. This perception of high impact may overlook future costs. As one participant cautioned, \textit{``short-term efficiency gains from \glsIfFirst{ai}{ }{-}generated solutions must be weighed against long-term maintainability''}, predicting that \textit{``the intermediate steps will cause pain''} (P207).

\section{Implications for Practice}
The empirical analysis presented highlights a rapid development in the ever-changing craft of software development. Similar shifts have happened before, albeit never as disruptive, e.g., when software engineering became its own discipline during the \enquote{software crisis}, when agile methods became popular in the early 2000s, or when DevOps and cloud strategies became omnipresent. Now, \gls{ai} tools have accelerated code generation but have also introduced new friction points regarding context, trust, and validation. Our findings suggest that mere adoption is no longer a differentiator, the real advantage lies in effective integration. 
Based on the presented results and key insights, we outline strategic implications for individuals, organizations, and tool vendors. 

\subsection{Individual Developers}
Our results indicate that individual productivity relies more on context engineering than on prompt engineering alone. As discussed in Section~\ref{sec:prompting_strategies}, strategies that reflect effective human-to-human delegation, such as providing clear context and specific instructions, are perceived as more effective than approaches like role prompting or predefined templates. Therefore, developers should focus on curating the context they provide to the model rather than memorizing prompt patterns.
The \secref{sec:communication_dividend} highlights a key shift in practice. Treating \gls{ai} as a collaborative partner that requires precise scoping and clarification yields better outcomes than treating it as an oracle. %

Our usage data reveals an asymmetry between generative and evaluative tasks. Code generation is the most common use case, while more specific activities such as bug fixing and test writing are less frequent. As syntax generation becomes increasingly commoditized, the core competency for software engineers is shifting toward rigorous code review and architectural reasoning. In addition, the correlations found between distrust and workflow speed highlight the cognitive cost of validation. To fully realize the benefits of \gls{ai}, developers should focus on strengthening their debugging and reasoning skills.

\subsection{Teams and Organizations}
Our findings show that organizations face challenges that go beyond the use of individual tools. The \secref{sec:experience-paradox} reveals a major gap between how junior and senior engineers assess the value and effectiveness of \gls{ai} tools. The \secref{sec:proficiency-cycle} indicates that the benefits of \gls{ai} adoption are not evenly distributed, with power users more commonly found in smaller organizations. The \secref{sec:corp-infra} demonstrates that larger organizations shape both the selection of tools and the ways in which they are used. Organizational scale influences not only which tools are deployed but also how they are integrated into workflows. In our sample, smaller enterprises adopt code-generation tools more widely than larger organizations, which may enable them to realize \glsIfFirst{ai}{ }{-}driven efficiencies more quickly.

Overall, these patterns indicate that successful \gls{ai} integration also relies on organizational capabilities rather than individual skills alone. The experience gap between junior and senior engineers can introduce challenges for collaboration and validation within teams. For instance, friction may occur when junior engineers produce \glsIfFirst{ai}{ }{-}assisted output that must be reviewed by more cautious senior colleagues, a dynamic highlighted by our interviewees. To address these issues, organizations should establish a structured knowledge exchange that encourages contributions from both senior and junior engineers. Senior engineers can share established heuristics and reasoning processes, while junior engineers can offer new strategies for interacting with \gls{ai} tools. Additionally, organizations should treat \glsIfFirst{ai}{ }{-}related knowledge, including context, prompt design, and validation methods, as a shared resource rather than an individual skill. This approach can help distribute the benefits of \gls{ai} adoption more broadly across the workforce.

\subsection{Tool Vendors}
Our analysis identified the \secref{sec:context-wall} as the most significant limitation. This limitation appears in two forms. The first is spatial context blindness, in which \gls{ai} systems struggle to understand the overall structure of a codebase. More than half of the respondents reported this as a major challenge. The second is temporal context decay, which results from knowledge cutoffs in pre-trained models and requires developers to manually update outdated implementations. According to our analysis of integration needs (Section~\ref{sec:integration}), understanding the full project context is the highest priority for customization, ranking above IDE integration and company-specific coding standards.

These findings indicate that the main challenge has shifted from improving model capability to strengthening contextual grounding. Therefore, tool vendors should focus on developing features that support guided exploration, allowing developers to direct the agent's attention to important architectural details. Interfaces should not rely solely on automatic context gathering, but should also enable user-driven context specification. In addition, effective tools need to balance autonomy with collaboration, since agents can access only information in the codebase and cannot infer requirements that are not specified. It is important to include features that clarify user intent before code generation. Implementing a structured planning process in which the agent and engineer jointly define the task could help ensure the model fully understands the context before generating code.

\section{Threats to Validity}
\paragraph{Internal Validity}
All values were measured via self-report, introducing measurement error due to imperfect recall or subjective estimation.
The fixed, thematically grouped question order may have introduced context or carryover effects, in which responses to earlier questions influenced subsequent answers.

\paragraph{External validity}
The participants were recruited via personal contacts and a LinkedIn post, resulting in a convenience sample that may not fully represent the broader user population. Also, most participants (89\%) were from Germany, as intended by design.
The findings were sampled in mid-2025 and may not generalize to future states of tool usage or workflows.
\paragraph{Construct validity}
The measurement of the tool usage frequency, the effectiveness of prompting strategies, and the impact on workflow speed rely on respondents' subjective interpretations and recall, introducing potential measurement error and limiting construct validity.

\section{Related Work}
Recent research has increasingly focused on integrating \gls{genai} into software engineering, especially following the introduction of tools such as GitHub Copilot and ChatGPT. In this section, we organize related work into two primary areas: developer-\gls{ai} interaction patterns and empirical studies on the adoption of \gls{ai} tools.

\subsection{Interaction Patterns and Productivity Effects}
Understanding how developers interact with \gls{ai} assistants has emerged as a critical research direction. \citet{barkeGroundedCopilotHow2022a} established that developer-AI interactions are bimodal: acceleration in which developers use \gls{ai} to achieve known goals faster, and an \textit{exploration mode} in which developers investigate options under uncertainty. 
Subsequent observational research by \citet{khojahCodeGenerationObservational2024a} revealed that practitioners more often use ChatGPT for guidance and learning (62\% of dialogues) rather than expecting ready-to-use code. A large-scale survey of 410 developers by \citet{liangLargeScaleSurveyUsability2024b} found that \gls{ai} assistants are primarily used to reduce keystrokes and recall syntax, but notably less for creative problem-solving. More recently, \citet{zakharovAISoftwareEngineering2025} investigated developers' mental models, identifying two primary framings: \Gls{ai} as a \textit{tool} versus \gls{ai} as a \textit{human-like teammate}. Assigning multiple roles to \gls{ai} correlated positively with perceived usefulness.

\citet{tafreshipour2025prompting} investigated the evolution of 1262 prompt templates in 243 GitHub repositories. They found that, while prompt changes are rarely documented, they often lead to inconsistencies or decreased performance of the \gls{ai}.

Controlled experiments have quantified productivity effects. In their study establishing productivity baselines, \citet{pengImpactAIDeveloper2023a} conducted a randomized experiment with 95 developers, finding the Copilot group completed tasks 56\% faster. Building on this work, \citet{cuiEffectsGenerativeAI} extended the analysis with field experiments across 4,867 developers at Microsoft, Accenture, and a Fortune 100 company, demonstrating a 26\% increase in completed tasks. Critically, these studies reveal substantial experience effects: junior developers showed 21-40\% productivity gains compared to 7-16\% for seniors. A recent counterpoint can be found in  \citet{beckerMeasuringImpactEarly20252025}, who found that experienced open-source developers using \gls{ai} tools actually increased completion time by 19\%, suggesting that expertise moderates \gls{ai} effectiveness in complex ways.

\subsection{Empirical Studies on \glsentrytext{ai} Tool Adoption}
The research field of \gls{ai} for supporting software engineering spans the whole software development life cycle. \citet{durrani2025impact} show that \gls{ai} tools are driving gains in accuracy and efficiency across most phases. While there is research on requirements engineering~\citep{Krishna2024Using,Marques2024Using}, software design and architecture~\citep{schmid2025softwarearchitecturemeetsllms}, and operations~\citep{Ye2025LLMSecConfig}, the most mature fields with many available tools and techniques are implementation~\citep{Jiang2024A} and quality assurance~\citep{Wang2023Software}.

AI-driven tools are transforming the software development life cycle, with applications spanning code generation, defect prediction, and automated testing~\citep{alenezi2025aidriven}. Multiple industry studies reveal rapid adoption of \gls{ai} tools, some as early as 2023~\citep{pashchenko2023early}, and more in the following years~\citep{chatterjee2024impactaitoolengineering,davila2024industry,russo2024navigating,li2024aitooluseadoption,lambiase2025exploringindividualfactorsadoption,lambiase2025investigating}. These studies report high adoption rates, especially for tools that integrate well into existing processes and developer habits. Enterprise-specific challenges are highlighted by \citet{weiszExaminingUseImpact2025}, who studied IBM's deployment of an \gls{ai} coding assistant and found that 43\% of users felt \textit{less} effective with the tool, pointing to barriers including compliance concerns and organizational culture. \citet{strayHumanAICollaborationSoftware2025} report similar findings in a public sector context, with workflow compatibility and experience level as key adoption factors.

The annual Stack Overflow Developer Survey\footnote{\url{https://survey.stackoverflow.co/}} shows active \gls{ai} usage increased from 44\% (2023) to 62\% (2024), with trust declining from 77\% to 60\% over the same period. Similarly, the JetBrains State of Developer Ecosystem\footnote{\url{https://www.jetbrains.com/lp/devecosystem-2025/}} survey of 24,534 developers found that 85\% regularly use \gls{ai} tools. Regional variation is documented in a GitHub survey on \gls{ai} in software development\footnote{\url{https://github.blog/news-insights/research/survey-ai-wave-grows/}}: 88\% of US organizations support \gls{ai} tool adoption versus only 59\% in Germany. Analysis of the data directly confirms this trend, showing high adoption rates in countries such as India, Brazil, and Ukraine, and lower adoption rates in more developed countries like the United States and Germany.

Different from previous studies, our work seeks to address gaps in understanding the depth of practitioners' interactions with \gls{ai} tools. While earlier surveys mainly report on tool usage and frequency, we systematically examine the prompting strategies practitioners employ, the specific challenges they face, and the influence of experience and organizational context on effectiveness. Our study is conducted in Germany, a market characterized by strict GDPR compliance requirements, a strong presence of small and medium-sized enterprises, and established engineering quality standards. By focusing on this context, we identify friction points that may not be as apparent in other markets.

\section{Conclusion}
In this study, we move beyond reporting adoption rates to analyze how German software engineers use \gls{genai} tools in their daily work. While prior research has shown that adoption is widespread, there is limited understanding of engineers' interactions with these tools, including prompting strategies, the influence of organizational constraints, and the role of experience in productivity. Our mixed-methods analysis identifies several key findings. First, we observe that junior and senior engineers perceive the effectiveness of \gls{ai} tools differently, which has implications for team collaboration. Second, we find that spatial and temporal context constraints pose significant barriers, requiring additional verification work. Third, our results indicate that the benefits of \gls{genai} tools are concentrated among experienced users, rather than being evenly distributed across all users. We also find that advanced prompting strategies are not commonly used in practice and that providing clear, direct context is generally more effective. Overall, our findings suggest that the value of \gls{genai} in organizations depends on effective integration into workflows and practical usage.

There are several directions for future research. First, addressing the limitations of our cross-sectional design through longitudinal studies would help establish causal relationships and track changes in interaction patterns as \gls{genai} tools evolve. Further, research should focus on practitioners' practical experiences, as our analysis shows that engineers often rely on straightforward strategies and encounter challenges that new tool features may not address. Finally, ongoing surveys are necessary to capture real-world practices and to identify challenges that may not be apparent in controlled settings.

\balance

\bibliographystyle{ACM-Reference-Format}
\bibliography{references}

\end{document}